\documentclass[useAMS,usegraphicx,usenatbib]{mn2e}

\usepackage{graphicx}
\usepackage{times}
\usepackage{amssymb}
\usepackage{amsmath}
\newif\ifAMStwofonts
\AMStwofontstrue

\def\pmn{{PMN~J0525--3343}}
\def\gb{{GB~1428+4217}}
\def\rx{{RX~J1028.6--0844}}

\def\asca{{\it ASCA}}
\def\xmm{{\it XMM-Newton}}
\def\bepposax{{\it BeppoSAX}}
\def\rosat{{\it ROSAT}}

\def\mhz{{\rm\thinspace MHz}}
\def\ghz{{\rm\thinspace GHz}}
\def\a{{\rm\thinspace\AA}}

\def\kmps{\hbox{$\rm\thinspace km~s^{-1}$}}
\def\pcmsq{\hbox{$\rm\thinspace cm^{-2}$}}

\def\ev{{\rm\thinspace eV}}
\def\kev{{\rm\thinspace keV}}

\def\ergpcmsqps{\hbox{$\rm\thinspace erg~cm^{-2}~s^{-1}$}}
\def\ergps{\hbox{$\rm\thinspace erg~s^{-1}$}}
\def\ergcmps{\hbox{$\rm\thinspace erg~cm~s^{-1}$}}

\title{\xmm\ confirmation of a warm absorber in the $z=4.4$ blazar \pmn}
\author[M. A. Worsley et al.]
{\parbox[]{6.in}
{M.~A. Worsley$^{1}$\thanks{E-mail: maw@ast.cam.ac.uk}, A.~C. Fabian$^{1}$, A.~K. Turner$^{1}$, A. Celotti$^{2}$ and K. Iwasawa$^{1}$}\\
\footnotesize
$^{1}$Institute of Astronomy, Madingley Road, Cambridge CB3 0HA\\
$^{2}$S.~I.~S.~S.~A., via Beirut 2-4, I-34014 Trieste, Italy\\
}

\begin{document}
\maketitle

\label{firstpage}

\begin{abstract}
We present the analysis and results from a series of \xmm\ observations of the distant blazar \pmn\, at a redshift of 4.4. The X-ray spectrum shows a spectral flattening below $\sim1\kev$ confirming earlier results from \asca\ and \bepposax. The spectrum is well fitted by a power-law continuum with either a spectral break or absorption; no sharp features are apparent in the spectrum. No variability is seen in the individual lightcurves although there is evidence of small longer term variations ($\sim$\thinspace months in the blazar frame) in both flux and $(0.1-2.4)/(2-10)\kev$ flux ratio. Very low levels of optical-UV extinction and the lack of any evidence of a Lyman-limit system at the quasar redshift rule out neutral absorption and we argue that the most plausible explanation is the presence of a warm, ionized absorber. Very strong C\thinspace\textsc{iv} absorption in the optical spectrum already implies the presence of highly ionized material along the line of sight. Warm absorber models using the photoionization code \textsc{cloudy} are consistent with both the X-ray and C\thinspace\textsc{iv} data, yielding a column density $\sim10^{22.3-22.5}\pcmsq$ and ionization parameter $\sim10^{0.8-1.2}\ergcmps$. 
\end{abstract}


\section{Introduction}
\pmn\ at $z=4.4$ is one the the most distant, X-ray bright, radio-loud quasars known \citep{fabian_pmn}. It is one of a handful of such objects that have characteristics typical of blazars with a spectral energy distribution showing peaks in the infra-red and $\gamma$-ray regimes, and displaying long-term variability \citep{fabian97,fabian98,moran97,zickgraf97,hook98}.

\pmn\ shows spectral flattening at soft X-ray energies, initially seen in observations made with \asca\ and \bepposax\ \citep{fabian_pmn} and now with \xmm . Above $2\kev$ the continuum is well fitted by a power-law but this either breaks or is absorbed below $\sim1\kev$. A similar effect has also been reported in the $z=4.72$ and $z=4.28$ blazars, \gb\ \citep{boller00,fabian_gb} and \rx\ \citep{yuan00}, respectively. If the soft X-ray deficit is attributed to cold absorption intrinsic to the objects then columns of $10^{22}-10^{23}\pcmsq$ are required. It was not clear, however, if the effect is due to absorption or an underlying break in the continuum.

Spectral flattening has been widely found in lower redshift radio-loud quasars. \citet{fiore98} found a systematic decrease with redshift of the spectral slope of soft X-ray emission from radio-loud quasars out to $z\sim3.9$. \citet{cappi97} reported evidence of absorption for $1.2<z<3.4$ radio-loud quasars using \asca\ observations and \citet{reeves00} also found systematic spectral flattening for \asca\ sources out to a redshift of 4.2.
 
Absorption has been suggested as the likely explanation but the data quality has been insufficient to distinguish between an intrinsic absorber at the source redshift or an absorption system along the line of sight. The latter is ruled out by statistical arguments (e.g. \citealp{o'flaherty97}). The lack of the effect in $z=2-3$ \emph{radio-quiet} quasars would tend to suggest an association with jet activity -- either due to an intrinsic absorber associated with the jet or an underlying continuum effect. If the flattening is due to intrinsic absorption there remains much speculation as to its nature. A number of the objects show low levels of UV-optical extinction that are inconsistent with such large columns of cold absorbing material and it has been suggested that the gas may be highly ionized \citep{fabian_pmn}, similar to the `warm absorber' as is common in many Seyfert galaxy nuclei (e.g. \citealp{reynolds97,crenshaw99}).

Recent \xmm\ observations \citep{grupe03} of the blazar \rx\ found only marginal evidence for excess absorption, much less than the high column density ($2\times10^{23}\pcmsq$; \citealp{yuan00}) implied by the \asca\ spectrum. We find that the soft X-ray flattening in \pmn\ and \gb\, as detected with \asca-\bepposax\ and \rosat-\bepposax, respectively, is reduced but robustly detected with \xmm\ observations (this work; Worsley et al., in preparation).

Alternatively, an underlying break in the continuum emission could explain the flattening; X-ray emission from powerful blazars is thought to be due to inverse Compton scattering by relativistic electrons in the jet plasma. The source photons for the scattering are supplied either by synchrotron emission from the same electron population, or, from an external radiation field of soft photons emitted from the nucleus. A spectral break can be explained in either mechanism by a low-energy cut-off in the electron population, or, in the second process, by a sharp peak in the spectrum of the soft photon field. The association of the break with the jet activity naturally explains the lack of the flattening in radio-quiet quasars. 

Here we present the results of a series of \xmm\ observations of \pmn. We confirm the spectral flattening as detected in previous \asca\ and \bepposax\ measurements \citep{fabian_pmn} and conclude that the most likely explanation is the presence of intrinsic warm absorption.

\section{Identification and the Optical Spectrum}

\pmn\ was identified by \citet{hook98} in a program to identify radio-loud quasars at high redshift. The quasar was selected as a flat spectrum radio source, with a spectral index $\alpha=-0.1$ in the Parkes-MIT-NRAO (PMN) Survey at $4850\mhz$ and the NRAO VLA Sky Survey (NVSS) at $1.4\ghz$ \citep{wright96,condon98}. X-ray detection was achieved with the \rosat\ High Resolution Imager on 1998 March 4--6, follow-up observations were made with \asca\ on 1999 March 1--2 and \bepposax\ on 2000 February 27--28.

\begin{figure}
\rotatebox{270}{
\resizebox{!}{\columnwidth}
{\includegraphics{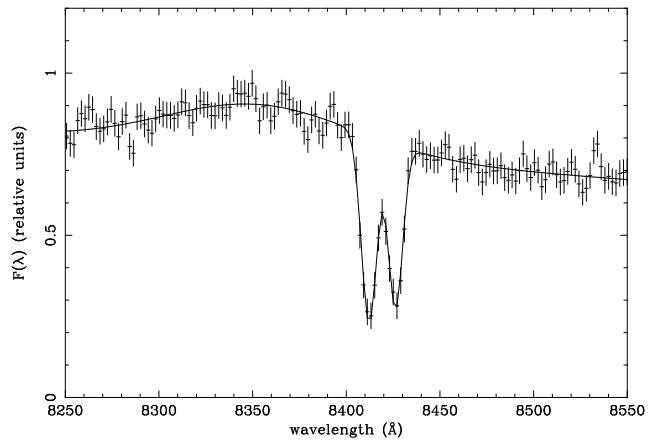}}}
\caption{Optical spectrum showing the peak of the C\thinspace\textsc{iv} emission line and redshifted absorption doublet. Produced from data obtained by \citet{peroux01}.}
\label{lines}
\end{figure}

\begin{figure}
\rotatebox{270}{
\resizebox{!}{\columnwidth}
{\includegraphics{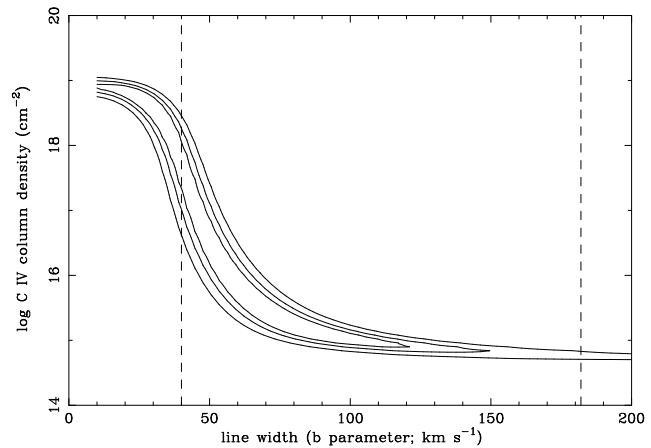}}}
\caption{Confidence plot of the C\thinspace\textsc{iv} column density and Doppler b parameter, 68, 90 and 99 per cent confidence regions are indicated. Doppler b parameter lower and upper limits are those implied by the Ly-$\alpha$ absorption line strength and the C\thinspace\textsc{iv} absorption line widths.}
\label{civ_conf}
\end{figure}

An optical spectrum in the range $4000-9000\a$ \citep{peroux01} is discussed by \citet{fabian_pmn}. Strong emission lines are seen due to the Ly-$\alpha$ + N\thinspace\textsc{v} blend along with broad ($11\thinspace500\kmps$ FWHM) emission due to C\thinspace\textsc{iv}. A deep C\thinspace\textsc{iv} absorption doublet ($1548.195$, $1550.770\a$) is seen, redshifted by $2700\kmps$ with respect to the peak of emission (Fig. \ref{lines}). Absorption is also seen due to Si\thinspace\textsc{iv}, N\thinspace\textsc{v} and Ly-$\alpha$. The absence of an optically-thick Lyman-limit system at the quasar redshift indicates that $N_{\rm{H\thinspace\textsc{i}}}<3\times10^{17}\pcmsq$. This constraint, together with the Lyman-$\alpha$ absorption line strength indicates the Doppler parameter, $b$, has to be at least $40\kmps$ \citep{fabian_pmn}. We have fitted here the C\thinspace\textsc{iv} absorption lines and find Doppler parameters of $b=156\pm26\kmps$ and $b=143\pm29\kmps$ respectively, suggesting an upper limit of $b\sim182\kmps$. The rest frame equivalent widths of $3.52\pm0.60\a$ and $3.16\pm0.65\a$ yield the column density in C\thinspace\textsc{iv} from a curve of growth analysis. Fig.~\ref{civ_conf} shows the confidence limits on the column density and $b$ parameter. Even given the upper and lower limits on the $b$ parameter the column is relatively unconstrained at $\sim10^{15-18}\pcmsq$. 

\section{X-ray Observations}

\xmm\ observed \pmn\ on 2001 February 11 and again on 2001 September 15. The first observation was badly affected by background flaring but the second obtained good data. These results indicated significant differences from the \asca-\bepposax\ spectra. This prompted us to apply for future observations to search for variability and a series of six more were made between 2003 February 14 and 2003 August 8. All observations were made in full frame imaging mode with the thin 1 filter.

The EPIC data reduction was performed with the Scientific Analysis Software (\textsc{sas}) \textsc{v5.4.1}. Periods of background flaring were excluded and the data were filtered in the standard way to include only the recommended event patterns for spectral analysis and to exclude events from `bad' pixels \citep{ehle03}. Table~\ref{exposures} shows the details of the observations. Background-corrected spectra were obtained from the observations, grouped to a minimum of 20 counts in each bin and analysed using \textsc{xspec v11.2} \citep{arnaud96}.

Spectral fitting to a simple power-law, modified only by galactic absorption ($N_{\rm{H}}=2.21\times10^{20}\pcmsq$; \citealp{elvis94}) was made over the full energy range of $0.3-12\kev$ to the data from the PN and MOS cameras simultaneously. The fit was poor (with a reduced chi-squared of $\chi^{2}_{\nu}=1.109$; $\nu=1913$) with highly curved residuals. Good fits, however, were made over the energy range $2-12\kev$. If the power-law photon index $\Gamma$ is allowed to vary in the $2-12\kev$ range between data sets then we have $\chi^{2}_{\nu}=1.013$ ($\nu=629$) with values of $\Gamma$ varying from $1.64$ to $1.74$. Requiring the same value of $\Gamma$ for all observations also results in a good $2-12\kev$ fit, with $\Gamma=1.67\pm0.03$ ($\chi^{2}_{\nu}=1.007$; $\nu=635$). 

\begin{table*}
\centering
\caption{Details of \xmm\ observations.}
\label{exposures}
\begin{tabular}{ccccccccc}
\hline
Revolution & Date & $2-10\kev$ Flux & \multicolumn{3}{c}{Good Exposure Times (s)} & \multicolumn{3}{c}{Total $0.3-12\kev$ Counts} \\
&& ($10^{-13}\ergpcmsqps$) & PN & MOS-1 & MOS-2 & PN & MOS-1 & MOS-2 \\
\hline
216 & 2001 Feb 11 & 7.0        & 2132  & 2890  & 2673      & 843  & 273  & 265  \\
324 & 2001 Sep 15 & 9.0        & 14602 & 18991 & 19986     & 5995 & 2399 & 2571 \\
\hline
583 & 2003 Feb 02 & 9.2        & 8532  & 11534 & 10871     & 3476 & 1482 & 1456 \\
588 & 2003 Feb 24 & 9.5        & 8263  & 10824 & 10528     & 3438 & 1414 & 1412 \\
593 & 2003 Mar 06 & 10.1       & 6953  & 9001  & 8795      & 3131 & 1237 & 1171 \\
598 & 2003 Mar 16 & 10.4       & 5639  & 6467  & 6269      & 2541 & 825  & 863  \\
603 & 2003 Mar 25 & 10.3       & 6104  & 5881  & 5877      & 3099 & 759  & 754  \\
671 & 2003 Aug 08 & 10.8       & 8039  & 9127  & 8522      & 4166 & 1521 & 1269 \\
\hline
\end{tabular}
\end{table*}
 
The background-corrected spectrum from revolution 216 was very poor, particularly in the softer energies, and was excluded from further analysis. The revolution 324 observation had a $2-10\kev$ flux of $9.0\times10^{-13}\ergpcmsqps$. During the last six observations the flux increases steadily from  $9.2-10.8\times10^{-13}\ergpcmsqps$ over a period of $\sim1\thinspace\rm{month}$ in the source frame (the error on the flux measurement in a single observation is $\sim0.4\times10^{-13}\ergpcmsqps$). For comparison the \asca\ $2-10\kev$ flux was $12\times10^{-13}\ergpcmsqps$. No statistically significant variability is present in any of the individual lightcurves.

The $0.1-2.4\kev$ flux increases from $5.4\times10^{-13}\ergpcmsqps$ (revolution 324) to $5.7-6.5\times10^{-13}\ergpcmsqps$ over the last six observations (flux errors are typically $\sim0.3\times10^{-13}\ergpcmsqps$). These can be compared to the \asca\ and \bepposax\ values of $5.2\times10^{-13}\ergpcmsqps$ and $3.3\times10^{-13}\ergpcmsqps$ respectively. The earlier \rosat\ count rate corresponds to a value of $4.2\times10^{-13}\ergpcmsqps$ (assuming the soft energy photon index measured by \asca-\bepposax).

\begin{figure}
\rotatebox{270}{
\resizebox{!}{\columnwidth}
{\includegraphics{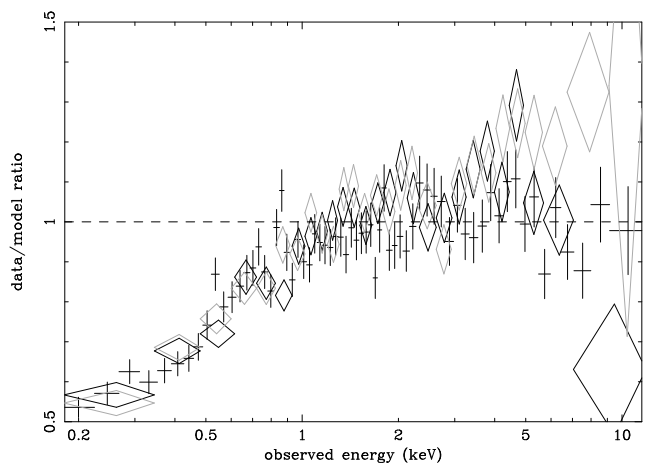}}}
\caption{Ratio of data to a simple $\Gamma=1.67$ power-law (including Galactic absorption of $N_{\rm{H}}=2.21\times10^{20}\pcmsq$; fitted over the $2-12\kev$ range), clearly showing the spectral break. The points are the combined spectra from revolutions 324--671 showing the PN (black crosses), MOS-1 (black diamonds) and MOS-2 (grey diamonds).}
\label{flattening}
\end{figure}

\begin{table*}
\centering
\caption{Results of the spectral fits. The 1-sigma error is shown for one parameter. The fits were made jointly to the PN and MOS data from all observations, over the range $0.3-12\kev$. Source fluxes were allowed to vary between observations but all other model parameters were fitted jointly.}
\label{fits}
\begin{tabular}{cccccccc}
\hline
Model & local $N_{\rm{H}}$ & intrinsic $N_{\rm{H}}$ & $\Gamma_{1}$ & $E_{\rm{break}}$ & $\Gamma_{2}$ & $\chi^{2}_{\nu}$ & $\nu$ \\
& ($10^{20}\pcmsq$) & ($10^{20}\pcmsq$) && ($\kev$) & & \\
\hline
Simple power-law & $2.21$ (Galactic) & $0$ & & & $1.53\pm0.02$ & $1.109$ & 1913 \\
\hline
Broken power-law & $2.21$ (Galactic) & $0$ & $1.2\pm0.2$ & $0.84\pm0.07$ & $1.61\pm0.03$ & $1.031 $ & 1911 \\
power-law, free local absorption & $5.9\pm0.4$ & $0$ & & & $1.67\pm0.02$ & $1.026$ & 1912 \\
power-law, free intrinsic absorption & $2.21$ (Galactic) & $134\pm13$ & & & $1.64\pm0.02$ & $1.024$ & 1912 \\
\hline
\end{tabular}
\end{table*}

\begin{figure}
\rotatebox{270}{
\resizebox{!}{\columnwidth}
{\includegraphics{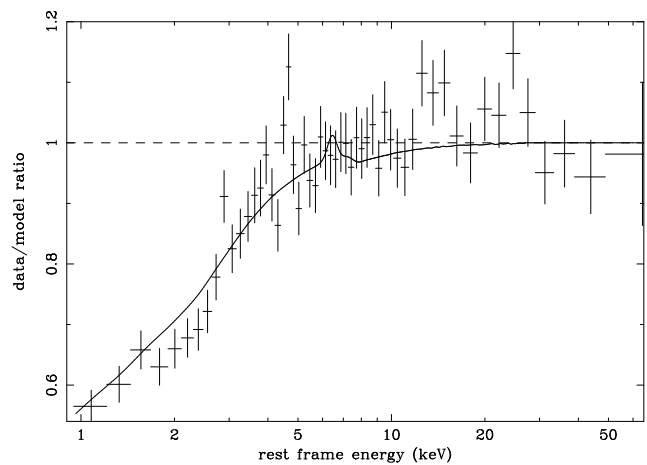}}}
\caption{The combined PN camera data of revolutions 324--671. The spectrum is plotted as a ratio to the $2-12\kev$ fitted power-law (inclusive of Galactic absorption). The energies are adjusted to the rest frame of the blazar (the observed frame range covered is identical to that in figure~\ref{flattening}). Overlaid is the most consistent warm absorption model prediction ($\xi=10\ergcmps$; $N_{\rm{H}}=10^{22.4}\pcmsq$). Fe K$\alpha$ emission is responsible for the feature at $6.4\kev$. It will be negligible if the warm absorber subtends a small solid angle at the source. The drop predicted due to absorption by iron is at $7.5\kev$ with an optical depth $\tau\sim0.02$.}
\label{combined}
\end{figure}

Fig.~\ref{flattening} shows the ratio of data to the $\Gamma=1.67$ fitted power-law (including Galactic absorption) as observed. Spectral flattening becomes apparent at low energies, with a clear break at $\sim1\kev$ and the data/model ratio falling to $\sim0.5$ for the softest energies. There appears to be little or no variation in spectral shape between the observations so the PN data from the different revolutions were combined together in the hope of detecting any features in the spectrum. Fig.~\ref{combined} shows the results (in the rest energy frame of the blazar). A break from the power-law occurs sharply at just below $\sim5\kev$. An emission-line-like feature seems to stand out at $\sim4.5\kev$ with a possible dip and/or another peak at around $\sim4\kev$. None of these correspond to any known spectral features or instrumental artefacts and given the number of data points we do not consider them statistically significant. The MOS camera spectra (Fig.~\ref{flattening}) do not corroborate any potential features. Upper limits on the detection of iron K have been estimated using the absorption edge above $7.1\kev$. This has an optical depth of $<0.07$ (corresponding to a column density of $<10^{17.6}\pcmsq$) while the K$\alpha$ emission line ($6.40\kev$) has an equivalent width $<13\ev$.

We considered a number of different models to take account of the the soft X-ray spectral flattening. The results are summarised in Table~\ref{fits}. All three models (broken power-law, power-law plus free local absorption and power-law plus free intrinsic absorption) produced statistically acceptable fits with no obvious irregularities in the residuals. The best-fitting models have improvements in chi-squared of between 164 (intrinsic warm/cold absorption) and 152 (broken power-law) when compared to the simple power-law (over the $0.3-12\kev$) range.

The fits show significant variation from the \asca-\bepposax\ observations which found local/intrinsic total absorption columns of $20/1800\times10^{20}\pcmsq$, compared to the $5.9/130\times10^{20}\pcmsq$ found here. The broken power-law fits to the \asca-\bepposax\ spectra also point to a stronger soft energy flattening; they require a break at a higher energy ($1.72\kev$) and to a flatter photon index ($\sim0.9$). The ratio of the \asca-\bepposax\ $(2-10)/(0.1-2.4)\kev$ flux is $\sim2.3$ compared to $\sim1.7$ for the \xmm\ data. The \asca-\bepposax\ observations show a $2-10\kev$ photon index of $\Gamma=1.47-1.68$ (broken power-law model; 90 per cent confidence) which is consistent with that found here ($\Gamma=1.67$).

To summarise, there is no evidence for variation in the spectral slope of the hard X-ray emission from either the \xmm\ or earlier observations. The $2-10\kev$ flux would appear to vary on $\sim6\thinspace\rm{month}$ time-scales ($\sim1\thinspace\rm{month}$ in the blazar frame) with amplitudes $\sim10-20$ per cent. The $0.1-2.4\kev$ flux also varies but the soft X-ray deficit appears to have lessened in the 1/2 years ($\sim2/4\thinspace\rm{months}$ in the blazar frame) between the \bepposax-\xmm\ and \asca-\xmm\ observations, characterised by a drop in the intrinsic cold absorber column by an order of magnitude.

\section{Discussion}

The four main potential explanations of the soft X-ray deficit will be examined in turn. Only two of the four scenarios (a break in the continuum or a warm absorber) appear to be self-consistent, the excess intergalactic and cold intrinsic absorber models both appear to be contradicted by other results. 

\subsection{Break in the Underlying Power-Law}

An acceptable fit to the data can be obtained by simply allowing a break in the underlying power-law. The best fit was found with a break occurring at $E_{\rm{break}}=0.84\kev$ where the photon index hardens from $1.61$ to $1.2$ in the soft X-ray band.

The spectral energy distribution (SED) of blazars is composed of two components; beamed synchrotron emission (peaking in the infra-red to soft X-ray regime) and a high-energy component (peaking in the X-ray up to $\gamma$-ray regime), see e.g. \citet{fossati98}. The likely mechanism for the production of high energy photons is the inverse Comptonization of relatively soft `seed' photons, by relativistic electrons in the jet \citep{sikora94}. The two sources for the seed photons are either the locally produced synchrotron emission, or an intense photon field (typically optical/UV) external to the jet, probably radiated from the accretion disc and reprocessed by circumnuclear gas. For the brightest blazars (as in \pmn) the latter, the so-called EC (external Compton) mechanism, is thought to dominate over the former, the so-called SSC (synchrotron self-Compton) mechanism.

Spectral breaks have been observed before in X-rays: Concave spectra arise when the observed band spans the steep synchrotron cut-off ($\Gamma\gtrsim2-3$) before the flat inverse Compton component dominates with $\Gamma\sim1.5-1.8$. A convex spectrum occurs when the band is dominated by the synchrotron component, the synchrotron rise--peak--cutoff leading to a gradual steepening of the spectral slope (from as flat as $\Gamma\sim1.5$ through the peak at $\Gamma\sim2$ and into the cut-off with $\Gamma\gtrsim2-3$ or more). \pmn\ shows a convex spectrum with a \emph{very} flat soft X-ray slope ($\Gamma\sim1.2$). However, the high energy photon index ($\Gamma\sim1.6-1.7$) is still far too flat to be the synchrotron peak/cut-off. The sharply-flattening convex spectrum is not explicable through any known combination of blazar emission components.

As the hard X-rays observed are compatible with being due to inverse Compton emission, a flattening of the spectrum can be explained by a low-energy cut-off in the electron population of the jet. A cut-off can theoretically produce a maximum flattening to a photon index of up to $\Gamma\sim0$ \citep{fabian_pmn} and is therefore not inconsistent with the observations here. It would, however, be very difficult to reproduce such a sharp break (the break is very distinct and occurs over an energy span $<5\kev$ in the source frame) without a highly energy dependent (and unknown) re-acceleration process. Further difficulties arise because there is no reported evidence of any breaks in nearby radio-loud quasars. 

An alternative explanation (applicable only to the inverse Compton emission due to the EC scenario, as is believed to be the case here) requires the seed photon distribution to be sharply peaked at or decrease rapidly below a certain frequency (e.g. \citealp{ghisellini96}). This mechanism may only be important in the more distant (brighter) radio-loud quasars where the EC emission may dominate over the SSC, hence the lack of an observable break in nearby (weaker) object where the SSC emission could mask the EC spectral decline. As with the alternative mechanism it remains far from clear how to account for the sharpness of the break, even considering the fact that we only see EC photons scattered in a small range of angles. In summary, while we cannot eliminate the broken power-law possibility we consider it very unlikely.

\subsection{Excess Cold Galactic/Intergalactic Absorption}

The flattening can be accounted for by line of sight absorption by cold material with a column of approximately $N_{\rm{H}}\sim3.7\times10^{20}\pcmsq$ (in addition to the Galactic level). It is very unlikely that such a large column, localised in the direction of \pmn , could be attributable to some kind of Galactic feature, particularly given the large mass of material which would be implied. 

Intergalactic absorption systems remain a possibility, in particular damped Ly-$\alpha$ systems (DLAs), which can have neutral hydrogen columns of up to several times $10^{21}\pcmsq$. In order to reproduce the observed absorption the line of sight to the quasar would need to intercept one or more very high column density DLAs of which the chances are low \citep{o'flaherty97,zwaan99}. Such an explanation, although it cannot be ruled out, is undermined further by the fact that the soft X-ray flattening has been observed in several other high-$z$ quasars and has not been seen in $z=2-3$ radio-quiet quasars.

\subsection{Intrinsic Cold Absorption}

For a cold absorber to be viable it must be intrinsic to the source, perhaps arising as the result of some kind of outflow after the initial formation of the central black hole or just because the host galaxy is young and gas-rich. 

An intrinsic cold absorber at the source resulted in an acceptable fit to the X-ray data, requiring a column in neutral hydrogen of approximately $1.3\times10^{22}\pcmsq$ (Table \ref{fits}). Such a large amount of obscuring material is completely at odds with the optical observations which imply a column of $<3\times10^{17}\pcmsq$ (based upon the lack of a Lyman-limit absorption system; \citealp{fabian_pmn}). Supersolar metal abundances would appear to unable to account for the factor of $\sim10^{5}$ difference required for $N_{\rm{H\thinspace\textsc{i}}}$. Moreover, the dust in any such column of cold material would be expected to provide strong UV extinction, contrary to observation. 

\subsection{Warm Absorption}

A warm (ionized) absorber is able to reproduce both the high X-ray and low optical/UV column densities. Photoionized gas could produce significant photoelectric absorption in the soft X-ray regime. It is likely any dust present in the nuclear regions would have been sublimated by the high source flux or high temperatures, resulting in very low absorption in the optical/UV band.

The optical spectrum shows strong C\thinspace\textsc{iv} absorption and, whilst the nature of the absorption is unclear, it is still evidence for large quantities of highly ionized material. Indeed, strong C\thinspace\textsc{iv} features are common in the spectra of Seyfert galaxy nuclei that have X-ray warm absorbers \citep{crenshaw99}.

We constructed warm absorber spectral models using the photoionization code \textsc{cloudy} \citep{ferland98}. We varied the underlying power-law photon index $\Gamma$, the column density $N_{\rm{H}}$ and the ionization parameter $\xi=L_{2-10}/nR^{2}$ ($L_{2-10}$ is the inferred isotropic luminosity in the $2-10\kev$ rest frame of the blazar, $n$ and $R$ are the absorber density and the distance from the source respectively). 

A grid of spectra were calculated by varying the ionization parameter, column density and photon index. Solar abundances have been assumed. Assuming no spectral break intrinsic to the source the inferred isotropic luminosity, in the $2-10\kev$ blazar frame, is $L_{2-10}\sim1\times10^{47}\ergps$.

\begin{figure}
\rotatebox{270}{
\resizebox{!}{\columnwidth}
{\includegraphics{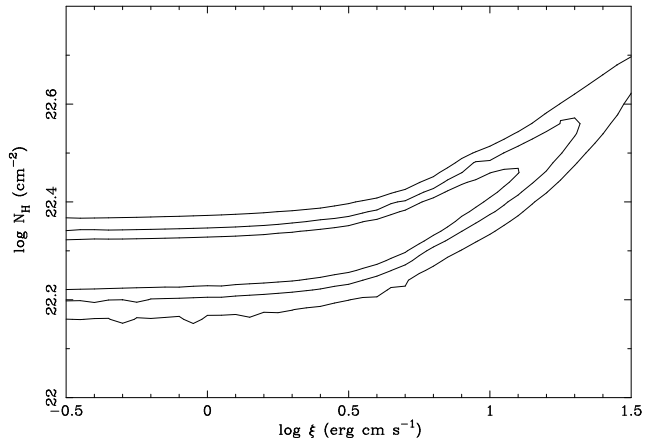}}}
\caption{Confidence plot for the ionization parameter and column density of the warm absorber model. 68, 90 and 99 per cent confidence contours are indicated.}
\label{confidence}
\end{figure}

Spectral fits were made to the X-ray spectra. Fig.~\ref{confidence} shows the confidence regions for the warm absorber ionization parameter and column density with a power-law index of $\Gamma=1.64$. The best fit is at $\log (\xi\thinspace/\ergcmps)=0.2$ and $\log (N_{\rm{H}}\thinspace/\pcmsq)=22.3$ but the confidence contours clearly show it was impossible to successfully constrain the parameters. The data is entirely consistent with a warm absorber scenario provided that $\log (\xi\thinspace/\ergcmps)\gtrsim0.8$ (for a column of $10^{22.3}\pcmsq$) so that the H\thinspace\textsc{i} edge has an optical depth $\tau<1$. In this case the warm absorber must have $\log~(\xi\thinspace/\ergcmps)\sim0.8-1.2$ and $\log~(N_{\rm{H}}\thinspace/\pcmsq)=22.3-22.5$. These models predict C\thinspace\textsc{iv} column densities $10^{14.5-18.4}\pcmsq$, consistent with the $\sim10^{15-18}\pcmsq$ implied by the equivalent width of C\thinspace\textsc{iv} line observed in the optical data. The fit of a $\xi=10\ergcmps$ $N_{\rm{H}}=10^{22.4}\pcmsq$ warm absorber model to the soft X-ray flattening is shown in Fig.~\ref{combined}. The mass of the absorber is $\sim3\times10^{9}\thinspace R^{2}_{\rm{kpc}}\Omega M_{\odot}$, where $\Omega$ is the solid angle subtended by the absorber at the source.

We note the UV absorbers seen in Seyfert galaxies are not the same matter giving rise to the X-ray absorption (e.g. \citealp{crenshaw03}), but are probably dense clouds. Higher resolution spectra of the UV absorption in \pmn\ is required to determine whether it too is due to separate clouds at different velocities, or to a smooth flow. 

It is unfortunate to note that the observed X-ray spectra contain no definite warm absorber signatures e.g. neutral or slightly ionized iron edges, which should occur at observed energies in the $1-2\kev$ range. Given the estimated column and ionization parameter the photoionization models do predict various absorption edges, principally due to oxygen and neon species, but these are in the $\sim0.1-0.25\kev$ range, redshifted out of the observed band. The continuum should recover below the oxygen absorption edge at $0.135\kev$ (our frame) which would be a convincing test of the warm absorber model. The MOS data are consistent with such a recovery but the calibration here is so uncertain that further work is required.

\section{Conclusions}

We have investigated the X-ray spectrum of the high redshift blazar \pmn\ with the \xmm\ observatory. We confirm the presence of significant soft X-ray spectral flattening, as previously detected in \rosat , \asca\ and \bepposax\ observations. The flattening can be explained by either excess absorption (either intergalactic or intrinsic to the source) or by an underlying break in the expected continuum emission (of a type not seen before). Optical observations, statistical arguments and the lack of any effect in radio-quiet quasars makes any kind of cold absorber seem unlikely -- the most plausible explanations remain either a highly ionized warm absorber at the source, or, a spectral break. 
 
A low energy cut-off in the emitting particle distribution responsible for the blazar inverse Compton emission could be a potential mechanism for the spectral break -- any such a cut-off would have significant importance in our knowledge of jet physics. The difficulty in this explanation is the lack of the effect in nearby radio-loud quasars, which show no evidence of a similar spectral break (which should be readily detectable at what would be energies $\sim5\kev$). An alternative, which would apply to the external Compton (EC) mechanism (as is believed to be important in very bright blazars such as \pmn), attributes the flattening to a highly peaked seed photon distribution. Neither process however, is able to easily account for the sharpness of the spectral break observed in \pmn\ which occurs over a source frame energy range of less than $5\kev$.

The presence of strong C\thinspace\textsc{iv} absorption, the lack of a Lyman-limit absorption system in the observed optical spectrum and low levels of UV reddening makes us favour a warm absorption model. Warm absorbers are a common feature of many observed Seyfert galaxies, where high ($\sim10^{21-23}\pcmsq$) column densities of highly ionized material reside close to the nucleus. We find the X-ray flattening is entirely consistent with a column $\sim10^{22.3-22.5}\pcmsq$ of gas with an ionization parameter $\sim10^{0.8-1.2}\ergcmps$. The equivalent width of the absorption implies C\thinspace\textsc{iv} column densities in the range $\sim10^{15-18}\pcmsq$, in agreement with the X-ray predicted $10^{14.5-18.4}\pcmsq$. A warm absorber also enables Ly-$\alpha$ and UV emission from the nucleus to be readily detectable. Further work on the C\thinspace\textsc{iv} absorber is required to refine the properties of the absorbing column.

\section{Acknowledgments}

Based on observations with \xmm, an ESA science mission with instruments and contributions directly funded by ESA Member States and the USA (NASA). MAW and AKT acknowledge support from PPARC. ACF thanks the Royal Society for support. AC thanks the Italian MIUR and ASI for financial support.

\bibliographystyle{mnras} 
\bibliography{mn-jour,MD1084rv}

\end{document}